\documentclass[pop,aps,floats,preprint,graphics,amsmath,amssymb,superscriptaddress,showpacs,epsfig,pre]{revtex4}

\usepackage{graphics,graphicx,epsfig}
\begin{document}
\title{Effect of current corrugations on the stability of the tearing mode.}

\author{F. Militello, M. Romanelli, R.J. Hastie, N.F. Loureiro \\
{\small EURATOM/UKAEA Fusion Association, Culham Science Centre,
Abingdon, Oxon, OX14 3DB, UK}}

\begin{abstract}

The generation of zonal magnetic fields in laboratory fusion
plasmas is predicted by theoretical and numerical models and was
recently observed experimentally. It is shown that the
modification of the current density gradient associated with such
corrugations can significantly affect the stability of the tearing
mode. A simple scaling law is derived that predicts the impact of
small stationary current corrugations on the stability parameter
$\Delta'$. The described destabilization mechanism can provide an
explanation for the trigger of the Neoclassical Tearing Mode (NTM)
in plasmas without significant MHD activity.

\end{abstract}

\maketitle

\section{Introduction}

In nature as well as in experimental devices, plasma turbulence
generates coherent structures, which can significantly affect the
overall behavior of the system. An example in magnetic fusion
plasmas is the occurrence of zonal electric fields or zonal flows
(the velocity is generated through $\mathbf{E}\times\mathbf{B}$
drifts). The zonal flows are predicted by electrostatic turbulence
theory \cite{Diamond2005} and are observed as mesoscale
oscillations in the velocity field, which fluctuates in the radial
direction while being homogeneous in the toroidal and poloidal
direction (i.e. $m=0$, $n=0$, where $m$ and $n$ are the toroidal
and poloidal wave numbers in a toroidal geometry). Such structures
are peculiar to turbulent systems since they are linearly stable
and can only exist as a result of nonlinear wave interaction.

The importance of the zonal flows as a self-regulating mechanism
for the turbulence and plasma transport is widely accepted. Only
recently, however, experimental observations
\cite{Fujisawa2007,Fujisawa2008} and nonlinear electromagnetic
simulations Ref.\cite{Thyagaraja2005,Waltz2006,Bruce} have
highlighted that turbulence in fusion plasmas can also generate,
together with the well documented zonal flows, zonal fields.
These, similar to the zonal flows, are axisymmetric sheared
band-like structures in the magnetic field. The zonal fields were
predicted by several theoretical works, a detailed review of which
can be found in Ref.\cite{Diamond2005}. In these works, zonal
magnetic fields are shown to arise from finite $\beta$ (the ratio
between the thermal and the magnetic pressure) drift-wave
turbulence when electron inertia effects are included.

The radial oscillation of the zonal magnetic field is associated
with zonal currents flowing in the direction of the confining
magnetic field. These turbulence generated \textit{current
corrugations} perturb the $m=0$, $n=0$ component of the current
density and its gradient, therefore affecting the stability of the
modes driven by it, such as the tearing mode \cite{FKR}. The
reconnection of the magnetic flux caused by the tearing
instability leads to the formation of so-called magnetic islands
which, if macroscopic, can significantly affect the performance of
fusion experimental device, as their particular topology enhances
the radial transport and therefore reduces the confinement.

The stability of the tearing mode is conveniently measured by the
stability parameter $\Delta'$, a positive value of which
corresponds, in the simplest theoretical models, to an unstable
mode \cite{FKR,Rutherford}. When pressure effects are neglected
and the boundary conditions of the problem are fixed, $\Delta'$ is
a function of the $m=0$, $n=0$ component of the current density
and of the wave vector of the perturbation, $\mathbf{k}$, only.

In Refs.\cite{Zakharov1989,Westerhof1990,Connor1992} it was shown
that, even if $\Delta'$ depends on the \textit{global} features of
the current density profile, it is strongly affected by the
\textit{local} structure of the current density at the resonant
surface where $\mathbf{k}\cdot \mathbf{B}=0$. This observation
suggested the possibility of controlling the tearing mode by
applying small localized current perturbations around the
reconnecting surface. In a similar way, however, turbulence
generated current corrugations can provide a source of free energy
for the tearing mode and modify its stability \cite{Adler1980}. As
a consequence, the zonal fields could be a possible triggering
mechanism for Neoclassical Tearing Modes, as they could push the
island above the seed limit. Implied in this reasoning is the
assumption that the corrugations are sufficiently coherent for the
tearing mode to have time to respond.

In this work we describe the effect of current corrugations of
given shape and amplitude on the calculation of $\Delta'$. By
using a simple slab model for the zonal fields we obtain an
analytical scaling for the variation of the stability parameter,
$\delta\Delta'$, with respect to the amplitude and the wave length
of the "zonal" flows that generate the current corrugations. We
then show numerically that the scaling applies also to more
relevant cylindrical cases. We find that even relatively small
current corrugations can induce large modifications of $\Delta'$
and therefore strongly affect the stability of the tearing mode.

A limitation of our work is the assumption, done for sake of
simplicity, that the pressure is flat in the island region.
Pressure gradients at the reconnecting surface, however, can play
an important role in the definition of the stability parameter, as
shown in Refs.\cite{FKR,Coppi1966}. We will address this subject,
together with a discussion of the generation of the "zonal"
pressure, in a future publication.

The paper is organized as follows. In Section II we describe the
basic model behind the calculation of $\Delta'$ for a given
equilibrium current density. We then describe the effect of small
modifications of the current density on the stability parameter by
using a perturbative approach. These results are applied to a slab
configuration in Section III, where a simple analytic scaling for
$\delta\Delta'$ is found. In Section IV we apply the scaling to a
cylindrical problem and we verify its validity using a numerical
code. Our results are used to investigate the onset of the NTMs in
the presence of current corrugations in Section V. Finally, in
Section VI we draw our conclusions.

\section{Model}

The 2D fluid equations employed in our analysis are well suited to
describe plasmas confined by a large toroidal magnetic field. We
restrict our attention to configurations with small inverse aspect
ratio, $\epsilon=a/R_0$, where $a$ and $R_0$ are the minor and
major radius of the machine. This leads to a simplification of the
geometry and, in particular, it allows the use of a cylindrical
tokamak approximation. In the following, the cylindrical
coordinates $(r,\theta,z)$ represent the radial, poloidal and
"toroidal" direction, respectively. Furthermore, we order $\beta$
as $\epsilon$, which implies that the electromagnetic effects are
relatively small, although not negligible.

We represent the magnetic field as:
\begin{equation}
\label{1} \textbf{B}=B_z \textbf{e}_z + \nabla \psi_p \times
\textbf{e}_z.
\end{equation}
where $B_z\cong const$ is the component of the magnetic field in
the "toroidal" direction (i.e. along the cylinder axis) and
$\psi_p$ is the poloidal magnetic flux. In order to describe a
single helicity tearing mode it is convenient to introduce an
helical coordinate: $\xi= \theta-\frac{n}{mR_0}z$. In the same
way, we express the magnetic flux in terms of its helicoidal
component, which is given by: $\psi=\psi_p+\frac{r^2nB_z}{2mR_0}$.
Finally, the helicoidal magnetic flux can be decomposed in its
equilibrium and perturbed part (the tearing mode) and the latter
can be expressed by a Fourier series:
$\psi=\psi_{eq}(r)+\displaystyle\sum_m \widetilde{\psi}_m(r)
e^{i(\gamma t - m \xi)}$. The magnetic flux and the "toroidal"
current density, $J$, are related through Ampere's law, so that
$(4\pi/c)J=-\nabla_{\perp}^2 \psi_p=-\nabla_{\perp}^2
\psi+\frac{2nB_z}{mR_0}$, where $\nabla_{\perp}^2
\psi=r^{-1}\partial_r (r\partial_r \psi)+ r^{-2}\partial^2_\xi
\psi$ and $c$ is the speed of light. The mode resonant surface,
where the flux reconnects, is such that the magnetic winding
number (the safety factor), $q(r)=\frac{rB_z}{R_0B_\theta}$, is
equal to $m/n$ ($B_\theta$ is the poloidal magnetic field). The
safety factor then can be easily expressed as a function of the
equilibrium "toroidal" current, $(4\pi/c)J_{eq}=-r^{-1}\partial_r
(r \partial_r\psi_{eq})+\frac{2nB_z}{mR_0}$:
\begin{equation}
\label{2} q=\frac{c}{4\pi}\frac{r^2B_z}{R_0\int dr [r J_{eq}(r)]}.
\end{equation}

In a high temperature plasma with low resistivity, and within our
approximations, the structure of the linear tearing mode far from
the resonant surface can be obtained by solving the scalar
equation $\mathbf{e}_z \cdot \nabla \times (\mathbf{J} \times
\mathbf{B}) =0$ (i.e. the projection of the curl of the plasma
momentum balance without inertial or viscous effects). Using
Eq.\ref{1} \cite{Strauss1976} and the definition of the helicoidal
flux it is possible to cast this equation in the following form:
\begin{equation}
\label{3} [J,\psi]=0,
\end{equation}
where the operator $[A,B]=r^{-1}(\partial_r A\partial_\theta B -
\partial_\theta A\partial_r B)$. The linear version of Eq.\ref{3}
is:
\begin{equation}
\label{4} \frac{1}{r}\frac{d}{d
r}\left(r\frac{d\widetilde{\psi}}{dr}\right)-\left(\frac{m^2}{r^2}-\frac{J_{eq}'}{\psi_{eq}'}\right)\widetilde{\psi}=0,
\end{equation}
where the helicoidal magnetic field is given by:
$-\psi_{eq}'=\frac{rB_z}{R_0}(\frac{1}{q}-\frac{n}{m})$, the prime
represents derivation with respect to $r$, the subscript $m$ is
dropped and the factor $4\pi/c$ is absorbed in the definition of
the current. It is clear now that the ideal MHD approximation
employed so far does not hold in the neighborhood of the resonant
surface, as the term proportional to $1/\psi_{eq}'$ in Eq.\ref{4}
yields a singularity.

The singular behavior is regularized by allowing for a small but
finite resistivity, which smooths out the current divergence
within a narrow boundary layer centered around the resonant
surface. As a consequence, we can identify two separated regions,
where different simplifications of the model equations apply: an
"outer" linear ideal region, where Eq.\ref{4} holds, and an
"inner" linear or nonlinear resistive boundary layer, the
narrowness of which allows geometrical simplifications, although
non-ideal physics must be retained. The solutions of the equations
in the two regions must match in a so called overlapping region.
In other words, the "outer" solution provides the boundary
conditions for the "inner" solution.

To simplify the matching procedure, it is convenient to introduce
the stability parameter, $\Delta'$:
\begin{equation}
\label{5}
\Delta'=\frac{1}{\widetilde{\psi}_{r_s}}\lim_{\epsilon\rightarrow
0}\left(\left.\frac{d \widetilde{\psi}}{d r}
\right|_{r=r_{s}+\epsilon} - \left.\frac{d \widetilde{\psi}}{d
r}\right|_{r=r_{s}-\epsilon} \right),
\end{equation}
where $\widetilde{\psi}_{r_s}$ is the value of the perturbed
magnetic flux at the resonant surface. Clearly, $\Delta'$ is fully
determined once the outer magnetic flux is found by solving
Eq.\ref{4}, and is a function of $m$, of $J_{eq}'/\psi_{eq}'$ and
of the boundary conditions imposed on $\widetilde{\psi}$. The
stability parameter defines the stability of the linear tearing
mode since, generally, a positive $\Delta'$ corresponds to an
unstable mode \cite{FKR}. Similarly, also in non-linear theory,
$\Delta'$ can be considered a measure of the stability of the mode
\cite{Rutherford,MP}.

We focus now on the modifications to the stability parameter due
to local changes of the profile of $J_{eq}'/\psi_{eq}'$. We first
assume a given equilibrium magnetic flux $\psi_{eq0}$, associated
with current density $J_{eq0}$ (and therefore to a safety factor
$q_0$). We then superimpose a "zonal" field, $\delta\psi_{eq}$,
which can be thought as the $m=0$, $n=0$ component of the magnetic
flux perturbation (and is accompanied by a "zonal" current $\delta
J_{eq}$ and a $\delta q$). The physical origin of the
corrugations, as discussed in the introduction, is suggested to
arise due to a combination of short scale turbulence and zonal
flows. We again emphasize that, in this calculation, we assume
that they are sufficiently consistent in time and space such that
a new "equilibrium" is temporarily created (hence the subscript
\textit{eq}). Obviously, there needs to be a constant source of
corrugations or otherwise they would dissipate on a time scale
proportional to $a^2/\eta$, where $a$ (of the order of the
collisionless skin depth) is the characteristic length scale of
the zonal field. In other words, $\delta\psi_{eq}$ directly
modifies $\psi_{eq0}$, thus generating the new "equilibrium":
$\psi_{eq}=\psi_{eq0}+\delta\psi_{eq}$ (together with
$J_{eq}=J_{eq0}+\delta J_{eq}$ and $q=q_{0}+\delta q$). This new
configuration affects the perturbed magnetic flux through
Eq.\ref{4} and therefore the stability parameter through
Eq.\ref{5}. In order to explicitly separate the effect of the
"zonal" field, we write the solution of Eq.\ref{4} as
$\widetilde{\psi} = \widetilde{\psi}_0 + \delta \widetilde{\psi}$,
where $\widetilde{\psi}_0$ is the eigenfunction associated with
the old equilibrium, while $\delta\widetilde{\psi}$ is the net
effect of the corrugations on the mode. Similarly, also the
stability parameter becomes: $\Delta' = \Delta'_0 + \delta
\Delta'$.

Following Refs.\cite{Zakharov1989,Westerhof1990}, we assume that
the localized current density corrugation is such that $\delta
\widetilde{\psi} \ll \widetilde{\psi}_0$, while $\delta
\widetilde{\psi}' \sim  \widetilde{\psi}'_0$. Furthermore, we
assume that $\delta q \ll q_0$, which is valid when $\int dr (r
\delta J_{eq}) \ll \int dr (r J_{eq0})$ (as in the case of a
localized zero-average oscillating $\delta J_{eq}$). With these
approximations [i.e. neglecting terms $O(\delta
\widetilde{\psi}/\widetilde{\psi}_0,\delta q/q_0)$] we obtain from
Eq.\ref{4}:
\begin{equation}
\label{6}
\frac{d}{dr}\left(r\frac{d}{dr}\delta\widetilde{\psi}\right)= -r
\frac{\delta J_{eq}'}{\psi_{eq}'}\widetilde{\psi}_0,
\end{equation}
and consequently, after integration:
\begin{eqnarray}
\label{6a} \left.\frac{d}{dr}\delta\widetilde{\psi}\right|_{r_s-\epsilon}&=& -\int^{r_s-\epsilon}_{0} dr\frac{r}{r_s-\epsilon} \frac{\delta J_{eq}'}{\psi_{eq}'}\widetilde{\psi}_0,\\
\label{6b}
\left.\frac{d}{dr}\delta\widetilde{\psi}\right|_{r_s+\epsilon}&=&\int^{\infty}_{r_s+\epsilon}
dr \frac{r}{r_s+\epsilon} \frac{\delta
J_{eq}'}{\psi_{eq}'}\widetilde{\psi}_0,
\end{eqnarray}
where we have employed the fact that any change in the perturbed
flux function must vanish far from the region where the current
corrugation is located (i.e. $d\delta \widetilde{\psi}/dr= 0$ both
for $r = 0$ and for $r = \infty$). From Eqs.\ref{5}, \ref{6a} and
\ref{6b} we obtain:
\begin{equation}
\label{7} \delta \Delta'=\frac{1}{r_s}\lim_{\epsilon\rightarrow
0}\left[\int_{0}^{r_s-\epsilon} rdr \left( \frac{\delta
J_{eq}'}{\psi_{eq}'}\frac{\widetilde{\psi}_0}{\widetilde{\psi}_{r_s}}\right)+\int_{r_s+\epsilon}^{\infty}
rdr \left( \frac{\delta
J_{eq}'}{\psi_{eq}'}\frac{\widetilde{\psi}_0}{\widetilde{\psi}_{r_s}}\right)\right]=
\frac{1}{r_s}P\int_{0}^{\infty} rdr \left( \frac{\delta
J_{eq}'}{\psi_{eq}'}\frac{\widetilde{\psi}_0}{\widetilde{\psi}_{r_s}}\right),
\end{equation}
where $P$ stands for the principal value of the integral
(necessary because the integrand is singular around the resonant
surface).

In the next Section we explicitly calculate $\delta\Delta'$ for a
given current density corrugation in slab geometry. This allows us
to obtain a semi-analytic scaling of $\delta \Delta'$ with respect
to the features of $\delta \psi_{eq}$.

\section{Slab case and $\delta \Delta'$ scaling}

In order to understand how a current corrugation affects the
stability parameter, we first investigate its effect in a simple
slab configuration. The problem is analyzed in a double periodic
rectangular box, where the normalized "radial" variable is $-\pi
\leq x \leq \pi$, and the normalized "helicoidal" variable is
$-\pi/k_y \leq y \leq \pi/k_y$, where $k_y$ is the aspect ratio of
the box (and also the wave number of the mode). We assume a given
normalized equilibrium current density, $J_{eq0}=\cos(x)$
[consequently, $\psi_{eq0}=\cos(x)$]. With this choice, resonant
surfaces form in the center of the box at $x=0$, and at the edge
at $x=\pm\pi$, where $\psi_{eq0}'=0$.

In this Section we study an elementary current corrugation,
$\delta J_{eq}= A K^2 \cos(Kx+\alpha)$, generated by a "zonal"
magnetic field \cite{Diamond2005} of given shape: $\delta
\psi_{eq}= A \cos(Kx+\alpha)$. Here $A$ is the (normalized)
amplitude of the "zonal" field, $K$ its wavelength, and $\alpha$
its phase with respect to the resonant surface. In order to
investigate its effect on the central resonance, we reduce
Eq.\ref{7} to a form that is appropriate to slab geometry:
\begin{equation}
\label{8} \delta \Delta'=P\int_{-\infty}^{\infty} dx \left(
\frac{\delta
J_{eq}'}{\psi_{eq0}'}\frac{\widetilde{\psi}_0}{\widetilde{\psi}_{r_s}}\right).
\end{equation}
Note that $x$ is normalized to a typical length scale and so is
$\delta \Delta'$. Furthermore, slab geometry requires symmetric
boundaries far from the reconnecting surface, so that the lower
extreme in Eq.\ref{7}, $r=0$,  is replaced by $x=-\infty$ in
Eq.\ref{8}.

For the equilibrium and the corrugation studied here, it is
straightforward to obtain:
\begin{equation}
\label{8a} \delta \Delta'= AK^3\left\{\cos(\alpha)
\int^{\pi/2}_{-\pi/2}dx\left[\frac{\sin(Kx)}{\sin(x)}\frac{\widetilde{\psi}_0}{\widetilde{\psi}_{r_s}}\right]+\sin(\alpha)
P
\int^{\pi/2}_{-\pi/2}dx\left[\frac{\cos(Kx)}{\sin(x)}\frac{\widetilde{\psi}_0}{\widetilde{\psi}_{r_s}}\right]\right\}.
\end{equation}
As a consequence of the parity of the integrand, the second term
on the right hand side of the previous equation is equals to zero.
At the same time, the first term on the right hand side does not
contain any singularity at the resonance, which implies that the
principal value notation can be dropped. Note also that the
integration limits are reduced to $x=\pm\pi/2$, which is halfway
between the edge and the central resonance but still
asymptotically far from the resistive layer (supposed to be
infinitesimally small). We restrict our attention to tearing modes
close to marginal stability, for which the constant-$\psi$
approximation \cite{FKR} is valid, i.e. $\widetilde{\psi}'_0 \ll
\widetilde{\psi}_0$ in the vicinity of the resonant surface. As a
consequence, for these modes we can assume that
$\widetilde{\psi}_0(x)/\widetilde{\psi}_{r_s} \cong 1$. Therefore,
the integral $\int^{\pi/2}_{-\pi/2}dx[\sin(Kx)/\sin(x)]$ can be
calculated for different values of $K$. For $K\geq 5$ its value is
roughly constant and can be approximated with $\pi$ (the maximum
error in this range is around $10\%$, and it reduces for large
$K$). We remark that the matching theory described here is valid
if $1/L \ll K \leq 1/\lambda$, where $L$ is a macroscopic length
scale and $\lambda$ is the resistive boundary layer width. Indeed,
if $K\lambda$ is too large the scale of the corrugation is of the
same order of the boundary layer, thus making impossible to employ
a perturbative technique, while if it is too small the effect of
the corrugations is evanescent.

To summarize, the modification of the stability parameter of a
constant-$\psi$ tearing mode depends linearly on the amplitude of
the "zonal" magnetic field and has a strong cubic dependence on
its wavelength :
\begin{equation}
\label{9} \delta \Delta' \approx C\cos(\alpha) A K^3,
\end{equation}
where $C$ is a constant depending on the equilibrium and the
system configuration (in the simple case treated here $C=\pi$).
When considering the features of the current corrugation, the
scaling is linear both in its wavelength ($K$) and amplitude
($AK^2$). Although this scaling was obtained with a basic model,
it applies also to more complicated geometries, as will be shown
in Section IV.

As an application, which is also useful to verify the validity of
Eq.\ref{9}, we study numerically the dispersion relation of a
tearing mode generated by a corrugated equilibrium. The
calculation is performed in the same 2D slab double periodic box,
equilibrium and corrugation described above. The equation that we
solve are the plasma vorticity equation and Ohm's law (resistive
reduced MHD model):
\begin{eqnarray}
\label{10} \frac{d U}{dt}&=&\nabla_{\|}J,\\
\label{11} \frac{d \psi}{dt}&=&S^{-1}(J_{eq}-J),
\end{eqnarray}
where all the quantities are normalized: the lengths with respect
to typical macroscopic scale, the time to the relevant Alfven
time, $S$ is the Lundquist number, and all the other variables
accordingly. The operator $d\cdots/dt=\partial\cdots/\partial t +
\mathbf{V}_{\mathbf{E}\times \mathbf{B}}\cdot\nabla\cdots$
contains the $\mathbf{E}\times \mathbf{B}$ drift velocity,
$\nabla_{\|}=\mathbf{B}/|\mathbf{B}|\cdot\nabla$ is the parallel
gradient, and the vorticity $U=\mathbf{e}_z\cdot\nabla\times
\mathbf{V}_{\mathbf{E}\times \mathbf{B}}$. The current density is
related to the magnetic flux through Ampere's law, which in a slab
gives: $J=-\nabla^2 \psi$.

The theoretical dispersion relation appropriate for this model is
\cite{Militello2003}:
\begin{equation}
\label{12} \gamma^{5/4}-\gamma^{1/4}b
S^{-1}=0.48(\Delta'_0+\delta\Delta') S^{-3/4}k_y^{1/2},
\end{equation}
where $b=[J_{eq0}''(0)+\delta J_{eq}''(0)]/[J_{eq0}(0)+\delta
J_{eq}(0)]$ (note that for $b=0$ we recover the classic solution
of Ref.\cite{FKR}, called FKR solution in the following).
Inspection of Eq.\ref{12} shows that current corrugation affects
the growth rate through both $\delta\Delta'$ and $b$.

The linearized version of the system \ref{10}-\ref{11} is solved
using a finite difference eigenvalue numerical code, benchmarked
with analytical cases. In all the simulations, the Lundquist
number is $S=10^{3}$. In absence of corrugations, the stability
parameter is uniquely defined by $k_y$ according to the relation:
$\Delta'_0=2(1-k_y^2)^{1/2}\tan[(1-k_y^2)^{1/2}\pi/2]$ (see e.g.
Ref.\cite{Grasso2001}). We start by choosing $k_y=0.94$, so that
the "smooth" equilibrium is characterized by $\Delta'_0=0.41$ and
it is therefore tearing mode unstable. Solving Eq.\ref{12}, we
find that the perturbation has growth rate $\gamma=0.0033$ (note
that the growth rate is normalized with respect to the Alfven
frequency) and the resistive layer has a width $\lambda\cong
0.045$.

\begin{figure}
\epsfig{file=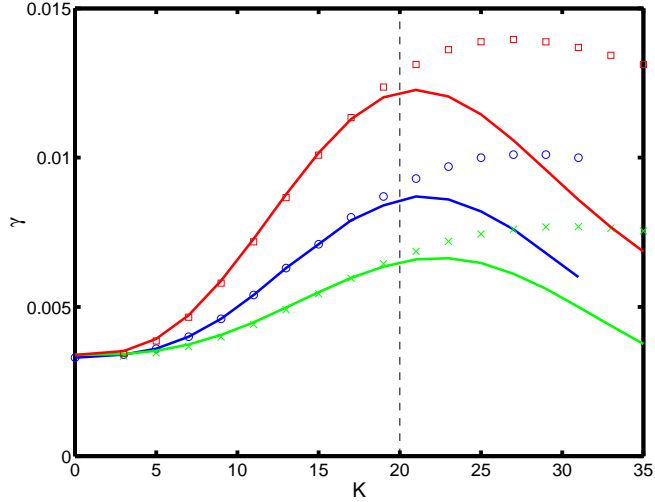, height=6.8cm} \caption{(Color online).
Growth rate of the tearing mode as a function of the corrugation
wavelength $K$. The solution is obtained for $S=10^3$,
$\Delta'_0=0.41$, $\lambda=0.045$ and $\alpha=0$. Crosses, circles
and squares represents cases with $A$ equals to $5\times 10^{-5}$,
$1\times 10^{-4}$, $2\times 10^{-4}$ respectively. The solid lines
show the theoretical prediction for the different cases. The thin
vertical line represents the limit of validity of the theoretical
prediction ($K\lambda=1$).} \label{fig3}
\end{figure}

The presence of a corrugation with wavelength $K$ can
significantly change the growth rate of the mode, as shown in
Fig.\ref{fig3}.
In the figure the crosses, circles and squares represent the
numerical solutions obtained with the linear code for corrugations
of amplitude $5\times 10^{-5}$, $1\times 10^{-4}$, $2\times
10^{-4}$, respectively. When $K<40$ (i.e. $K\lambda<1.82$), these
values lead to current corrugations of maximum amplitude smaller
than 7\%, 15\% and 30\% of the equilibrium current density. We
remark that even small amplitude corrugations can strongly modify
the growth rate of the mode. For example, in the case shown here a
corrugation that amounts to 4\% of the amplitude of the
equilibrium current density (e.g. $A=1\times 10^{-4}$ and
$K\lambda=0.91$) can lead to a growth rate three times as large
than in the "smooth" case. For values of $K\lambda$ smaller than 1
($K<20$ in our example), the theoretical predictions obtained from
the dispersion relation Eq.\ref{12} together with the scaling
Eq.\ref{9}, are in excellent agreement with the numerical data.
The disagreement for larger values of the corrugation's wavelength
is due to the limitations of the validity of the scaling.
Furthermore, also the dispersion relation Eq.\ref{12} is not
correct when the constant-$\widetilde{\psi}$ approximation does
not hold anymore.

\begin{figure}
\epsfig{file=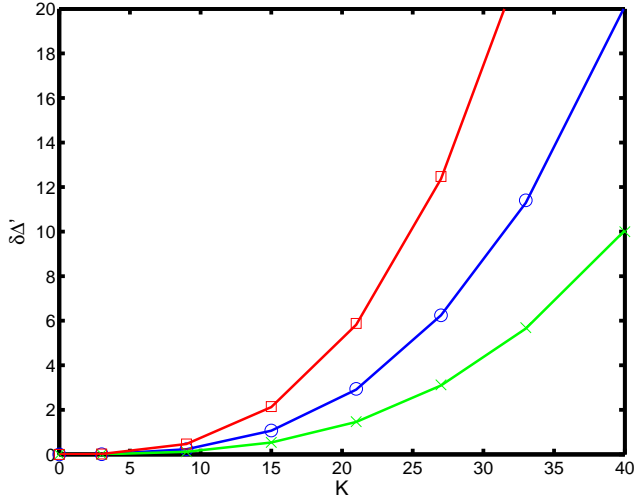, height=6.8cm} \caption{(Color online).
$\delta\Delta'$ as a function of the corrugation wavelength $K$,
for $\Delta'_0=0.41$ and $\alpha=0$. Crosses, circles and squares
represents cases with $A$ equals to $5\times 10^{-5}$, $1\times
10^{-4}$, $2\times 10^{-4}$ respectively. The solid lines show the
theoretical prediction (Eq.\ref{9}) for the different cases.}
\label{fig4}
\end{figure}

To complement this analysis, we have independently calculated with
a different numerical code the value of $\delta\Delta'$ associated
to the corrugations and the equilibrium described above. The code,
that is also used to obtain the cylindrical results presented in
the next Section, can solve Eq.\ref{4} or its slab version by
using a "shooting" algorithm. As Fig.\ref{fig4} shows, the
numerical values of $\delta\Delta'$ perfectly match the cubic
behavior predicted by the scaling in Eq.\ref{9}, thus confirming
its validity.

\begin{figure}
\epsfig{file=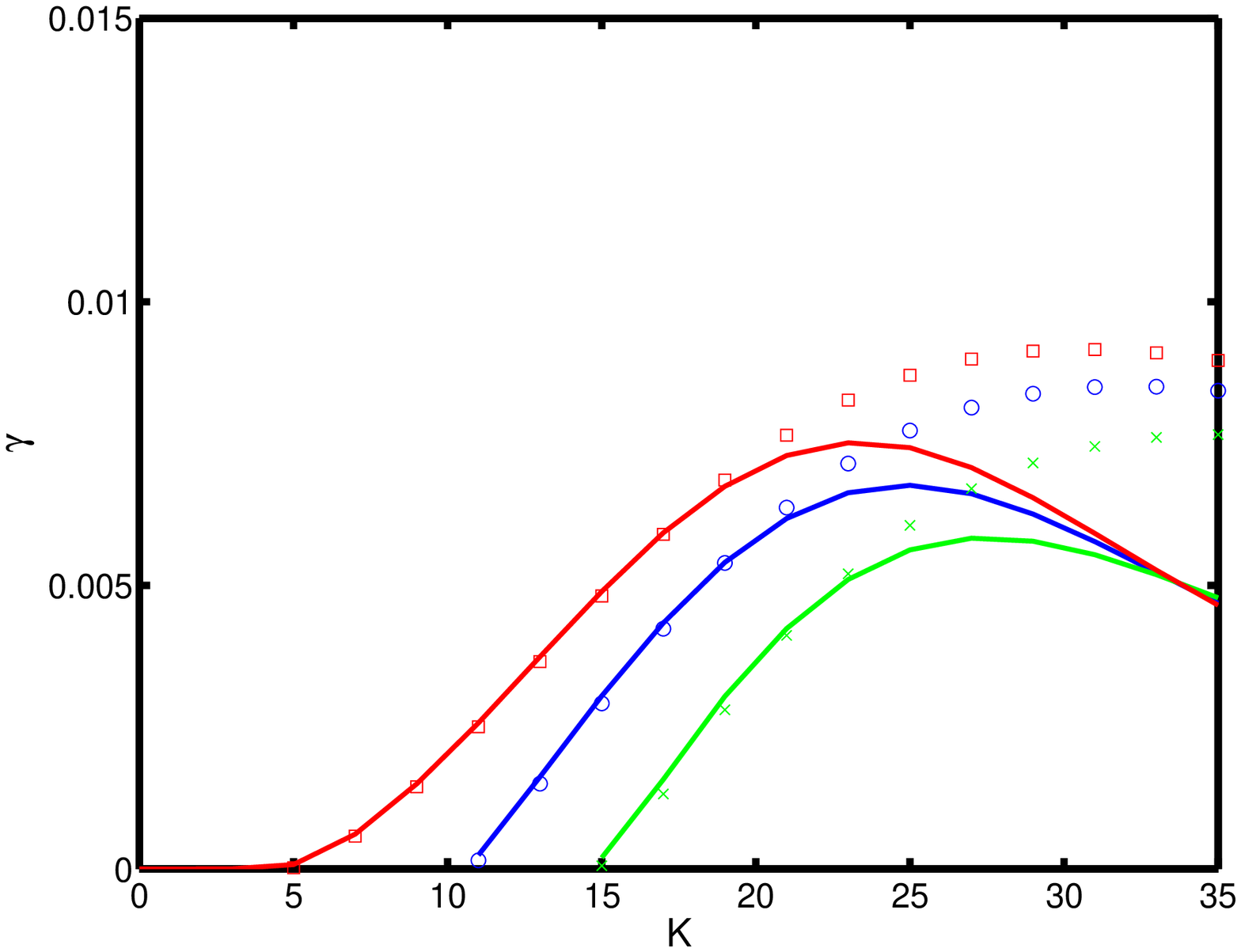, height=6.8cm} \caption{(Color online).
Growth rate of the tearing mode as a function of the corrugation
wavelength $K$. The solution is obtained for $S=10^3$, $A=1\times
10^{-4}$ and $\alpha=0$. Crosses, circles and squares represents
cases with $\Delta'_0$ equals to $-0.8$, $-0.29$ and $0$
respectively. The solid lines show the theoretical prediction for
the different cases.} \label{fig5}
\end{figure}

We complete this study by showing that a current corrugation with
the appropriate phase can drive a linear tearing mode even when
the "smooth" equilibrium would assure stability. In Fig.\ref{fig5}
we describe three cases, with $\Delta'_0=0$ (red squares),
$\Delta'_0=-0.29$ (blue circles) and $\Delta'_0=-0.8$ (green
crosses), for a corrugation of amplitude $A=1\times 10^{-4}$.
As expected, if the wavelength of the corrugation exceeds a
critical threshold, depending on $\Delta_0'$, the instability can
be excited. We conclude remarking that, for $\alpha=0$, the choice
of a positive $A$ leads to destabilization, while a negative value
would produce the opposite effect.

\section{Effect in cylindrical geometry}

The analytic results in Section III suggest the possibility that
the stability of the tearing mode can be significantly affected by
the presence of small scale, small amplitude current corrugations.
The scaling that we have obtained is verified by numerical codes
in slab geometry, and its validity is confirmed in this Section
for cylindrical geometry, which is more relevant for experimental
applications. In order to proceed, we perform a numerical study
with the "shooting" code introduced in the previous Section.

We assume a "rounded" equilibrium current density \cite{Furth1973}
given by: $J_{eq0}=J_0/[1+(r/r_0)^4]^{3/2}$, which implies an
equilibrium magnetic flux: $\psi_{eq0}=-0.25 J_0r_0^2\sinh^{-1}[
(r/r_0)^2]$. Here $J_0$ represents the current density at the
cylinder axis and $r_0$ is a measure of the width of the current
channel. In the following we take $J_0=2/0.7$ and
$r_0=\sqrt{1/5}$, so that the mode $m=2$, $n=1$ resonates at
$r_s\cong0.73$ and $m=3$, $n=2$ at $r_s\cong0.61$. In our model,
the "zonal" field that generates the current corrugation is
sinusoidal and localized around a radius $R$ by a Gaussian
envelope of width $\rho$ (taken as $0.2$ in all the simulations):
$\delta\psi_{eq}=A\cos[K(r-R)]e^{-(r-R)^2/\rho^2}$.

\begin{figure}
\epsfig{file=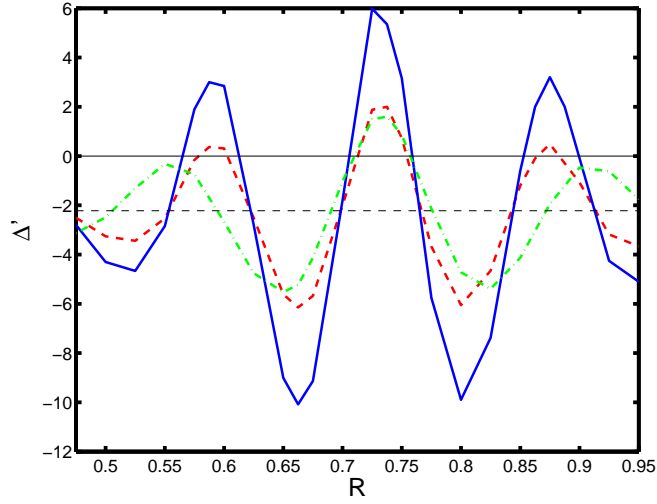, height=6.8cm} \caption{(Color online).
$\Delta'$ as a function of the localization radius $R$, for the
mode  $m=2$, $n=1$ and the equilibrium described in the text (for
which $\Delta'_{2,1}\cong-2.19$). The solid, dashed and dash-dot
lines describe cases with $A=5\times10^{-5}$ and $K=40$,
$A=2.5\times10^{-5}$ and $K=40$, and $A=5\times10^{-5}$ and
$K=30$, respectively.} \label{fig6}
\end{figure}

\begin{figure}
\epsfig{file=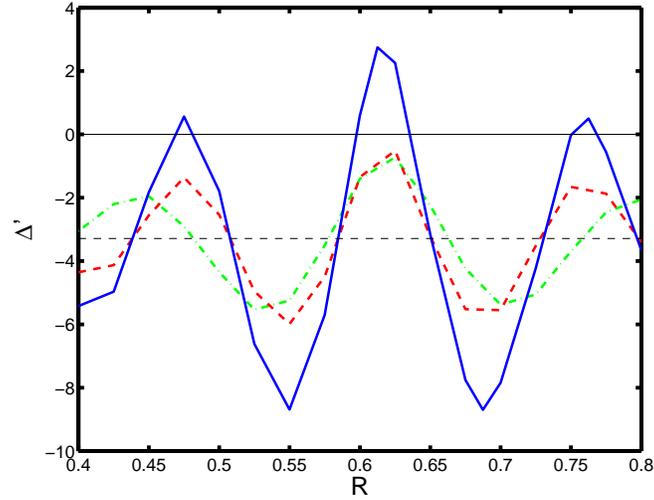, height=6.8cm} \caption{(Color online).
$\Delta'$ as a function of the localization radius $R$, for the
mode  $m=3$, $n=2$ and the equilibrium described in the text (for
which $\Delta'_{3,2}\cong-3.28$). The solid, dashed and dash-dot
lines describe cases with $A=5\times10^{-5}$ and $K=40$,
$A=2.5\times10^{-5}$ and $K=40$, and $A=5\times10^{-5}$ and
$K=30$, respectively.} \label{fig7}
\end{figure}

We investigate the effect of the corrugation on $m=2$, $n=1$ and
$m=3$, $n=2$ modes, which are stable in the "smooth" configuration
since $\Delta'_{2,1}\cong-2.19$ and $\Delta'_{3,2}\cong-3.28$. We
observe that the effect produced by $\delta\psi_{eq}$ on the
stability parameter changes as the localization radius of the
corrugation is modified and it reaches its maximum when $R\cong
r_s$ (see Figs.\ref{fig6} and \ref{fig7}).
This is expected, since the rigid shift around the resonance of
the sinusoidal perturbation is equivalent to a change in the phase
$\alpha$ in in the slab model. Furthermore, the comparison between
the solid ($A=5\times 10^{-5}$) and dashed lines ($A=2.5\times
10^{-5}$) in Figs.\ref{fig6} and \ref{fig7} confirm also in
cylindrical geometry the linear dependence of $\delta\Delta'$ on
the perturbation amplitude, as given in Eq.\ref{9}. Similarly,
also the scaling in the cube of the corrugation wavelength is
verified by measuring the maximum amplitude of the solid ($K=40$)
and the dash-dot ($K=30$) lines.

Finally, it is interesting to evaluate the constant $C$ for the
cylindrical equilibrium investigated. In order to do that, we
divide the maximum value of $\delta\Delta'$ (obtained at $R=r_s$)
by the amplitude of the corrugation and the cube of its
wavelength. With this procedure we find that $C\cong 2.5$ for
$m=2$, $n=1$ and $C\cong 1.7$ for $m=3$, $n=2$. While both values
are pretty close to the slab estimate, this result suggest a
dependence of the effect of the corrugation on the wave-vector of
the driven perturbation. This dependence is easily explained by
the presence of the $n/m$ term in the definition of $\psi_{eq0}'$
that, in cylindrical geometry, appears in the integral of
Eq.\ref{8}.

\section{NTM triggering mechanism}

In this Section, we analyze how the zonal fields affect the onset
of the Neoclassical Tearing Modes. In order to do that, we employ
a reduced version of the generalized Rutherford equation
\cite{Rutherford,LeHaye2006,Itoh2004}:
\begin{equation}
\label{13} \tau_\eta\frac{dw}{dt}\propto
\Delta_0'+\delta\Delta'+C_0\left(\frac{w}{w^2+w_{boot}^2}-\frac{ww_{pol}^2}{w^4+w_{pol}^4}\right),
\end{equation}
Consistently with Section III, the island width and the stability
parameter are normalized to a macroscopic length scale (e.g. the
resonant radius). For sake of simplicity, we retain in the model
only the fundamental terms responsible for the
stabilization/destabilization of the NTM. In particular, we
neglect the current shape term, that strongly affects the island
saturation but has little effect on the seed island problem
\cite{Militello2004,Escande2004} and the stabilizing toroidal
curvature effects \cite{Glasser1975}. Furthermore, we do not take
into account the rotation of the magnetic island in a
self-consistent manner \cite{CWW,Militello2008}. Indeed, in
Eq.\ref{13} neither the bootstrap term (third on the RHS), nor the
polarization term (fourth on the RHS) are explicit functions of
the island rotation. The coefficient $w_{boot}$, is a measure of
the reduction of the drive associated with the perpendicular
transport \cite{Fitzpatrick1995}, $w_{pol}$ is the cut-off due to
the banana orbit effect \cite{Bergmann2002} and $C_0$ is a
constant proportional to the poloidal $\beta$ \cite{LeHaye2006}.

We remark that our purpose is to sketch with an heuristic model
the mechanism that could allow NTM formation. A detailed analysis
would require a self consistent treatment of Rutherford equation
coupled with an equation for the rotation frequency of the
magnetic island. The theoretical uncertainties on the exact form
of this coupling make an attempt to employ more complicated models
to study the problem at hand a futile exercise.

In absence of zonal fields and if $\Delta_0'<0$, Eq.\ref{13}
predicts a minimum island width, $w_{seed}$, below which the
perturbation is always stabilized and the NTM does not reach
macroscopic size. Conversely, when the threshold is exceeded, the
dynamical evolution takes the island to its stationary saturation
amplitude, $w_{sat}$. To destabilize the NTM an island of
sufficient size is therefore required. This "seed" island could be
produced by some other MHD activity in the plasma, such as
sawteeth, fishbone instabilities or Edge Localized Modes. In the
following we investigate NTM dynamics without a seed [i.e.
$w(0)=0$] island but in presence of zonal fields.

From experimental observations \cite{Fujisawa2007,Fujisawa2008},
we expect that the zonal fields maintain their coherence for a
limited amount of time, $\tau_{zf}$. This fact is modelled by
assuming a temporal variation of the modified stability parameter,
so that $\delta\Delta'=\delta\Delta_{max}'\cos[(2\pi/\tau_{zf})
t]$, where $\delta\Delta_{max}'$ is a constant. A critical
parameter of the problem is therefore the ratio between this time
scale and that typical of the evolution of the NTM. From
Eq.\ref{13} we infer that this time is a fraction of the local
resistive time (calculated with the neoclassical resistivity and
the minor radius) $\tau_{NTM}\sim 0.1 \tau_\eta$, since the
saturated island width can be usually estimated around one tenth
of $r_s$. We expect that rapidly varying zonal fields, such that
$\zeta\equiv\tau_{zf}/\tau_\eta\ll 1$, will not be an efficient
trigger for the NTM as their effect is averaged to zero. On the
other hand, if $\tau_{zf}/\tau_\eta$ is of the order or greater
than one (although the latter possibility is difficult to achieve
in fusion plasmas) the zonal fields can drive $w$ above its seed
island limit.

\begin{figure}
\epsfig{file=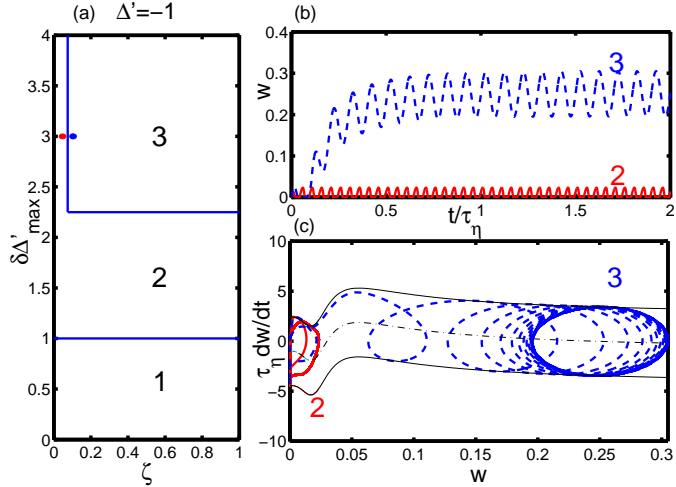, height=6.8cm} \caption{(Color online).
NTM triggering mechanism. (a) Approximate boundaries of the
relevant solutions of Eq.\ref{13}, obtained from the numerical
solution of Eq.\ref{13}. Parameters $\delta\Delta'$ and $\zeta$ in
zone 1 do not produce any island, in zone 2 yield to small
($w<w_{seed}$) oscillating islands, while in zone 3 islands reach
full NTM saturation. Physical parameters used described in the
text. (b) Time evolution and (c) phase diagram of an island with
parameters in zone 2 ($\zeta=0.05$, $\delta\Delta'_{max}=2.94$,
solid line) and in zone 3 ($\zeta=0.1$,
$\delta\Delta'_{max}=2.94$, dashed line). The thin solid lines in
(c) show $\tau_\eta dw/dt$ for the maximum and minimum value of
$\delta\Delta'$, while the dash-dot line for $\delta\Delta'=0$.}
\label{fig8}
\end{figure}
The solution of Eq.\ref{13} leads to three basic dynamic behaviors
for a stable configuration ($\Delta_0'<0$) in the presence of a
corrugation. When $\delta\Delta'_{max}<|\Delta'|$ the corrugation
is not strong enough to produce any island (region $1$ in
Fig.\ref{fig8}(a)). If the modified stability parameter is
sufficiently large, but smaller than a critical value
$\delta\Delta_{max}^{'crit}$, or
$\tau_{zf}/\tau_\eta<\zeta_{crit}$, an oscillating island appears
(region $2$ in Fig.\ref{fig8}(a)). Its maximum width remains below
$w_{seed}$ and is roughly proportional to $\tau_{zf}$. Finally,
when the critical values are exceeded, the full NTM is triggered
and the system settles in a state with $w$ oscillating around
$w_{sat}$ (region $3$ in Fig.\ref{fig8}(a)). The approximate
boundaries of these regimes [obtained solving Eq.\ref{13}
numerically] are sketched in Fig.\ref{fig8}(a), for a case with
$\Delta'=-1$, $C_0=0.25$, $w_{boot}=w_{pol}=0.025$, which
correspond to reasonable experimental values
\cite{LeHaye2006,Buttery2002}. For clarity, in Fig.\ref{fig8} (b)
we show the time evolution of two cases with
$\delta\Delta'_{max}=2.94$, one just below $\zeta_{crit}$ (
$\zeta=0.05$ for the solid line) and one above ($\zeta=0.1$ for
the dashed line). For the same two solutions, we plot in
Fig.\ref{fig8} (c) the phase diagram, which describes how $dw/dt$
and $w$ are related during the evolution of the perturbation. Two
dots in Fig.\ref{fig8}(a) display the position of these cases in
the $\delta\Delta'_{max}-\zeta$ space.

Obviously, in a realistic situation the time evolution of
$\delta\Delta'$ would be erratic and Eq.\ref{13} would become a
stochastic differential equation. We expect that this would make
the triggering mechanism easier, with a lower threshold for both
the critical $\zeta$ and $\delta\Delta'$. An interesting
discussion of the stochastic behavior of Rutherford equation was
given by Itoh \textit{et al.} in Ref.\cite{Itoh2004}. Furthermore,
the calculation presented here is conservative and the critical
values obtained are very stringent, since we have assumed an
initial condition for Eq.\ref{13} with no magnetic island. In
reality, the presence of a small, but finite seed island island
[$w(0)<w_{seed}$] would move the stability boundaries in
Fig.\ref{fig8}, broadening the unstable region 3.

\section{Conclusions}

We have investigated the effect of current corrugations on the
stability of the tearing mode. We have obtained a theoretical
scaling which relates the change of $\Delta'$ to the
characteristics of the corrugation, such as its amplitude, typical
scale length and phase with respect to the resonant surface. The
scaling has been tested on a simple slab linear problem, giving
excellent agreement between theoretical predictions and numerical
data. Then, by using a shooting code, we have investigated the
effect of the corrugations in cylindrical geometry, and found
qualitative and quantitative agreement with the analytic scaling.
Finally, we have addressed the problem of the seedless trigger of
a NTM through a current corrugation.

From our results we conclude that even small modifications of the
original equilibrium can lead to significantly different values of
the stability parameter and can therefore strongly affect the
onset and the dynamics of the mode. For example, a current
corrugation of amplitude around $5\%$ of the equilibrium and with
scale length comparable with the resistive layer can triple the
growth rate of the tearing instability. The study presented in
this paper has two clear implications for tokamak experiments.

First, it implies that the calculation of the stability parameters
from experimental data is an extremely delicate procedure. Indeed,
the current density profiles used to evaluate $\Delta'$ cannot be
directly measured. While the magnetic field can be deduced from
Motional Stark Effect measurements, the best spatial resolution
achievable is around 5cm, too coarse to rule out the possibility
of current corrugations. Another option is to reconstruct the
current profile from temperature and density data by using
equilibrium codes. However, the error bars of the measurement and
the approximations used in the current reconstruction make it
virtually impossible to obtain a precise value of the stability
parameter.

The second implication is related to the trigger of the NTMs in a
relatively quiescent plasma. Experiments in several machines
\cite{Gude1999,Buttery2002,Fredrickson2002} have reported
Neoclassical Tearing Modes growing without strong MHD precursors
(sawteeth, fishbones or ELMs) which could generate seed islands.
The theory presented in this paper suggests the possibility of
NTMs triggered by the microscopic turbulence via the generation of
slowly evolving (on the NTM time scale) current corrugations. The
corrugated configuration could be linearly unstable and generate a
seed island, which would then be sustained by Neoclassical
physics. At this point, the pressure flattening in the island
region could remove the drive for the turbulence and therefore for
the zonal field, which would decay, leaving the island at the NTM
saturated level. The heuristic model that we have employed shows
that a seedless NTM can indeed be excited if the zonal flow that
produces the corrugation is sufficiently strong and slowly
evolving. Furthermore, even when the zonal field is not able to
directly drive the NTM, it can significantly reduce the seed
island width (i.e. the plasma becomes more sensitive to the
presence of small islands).

It is useful here to give some experimental values for the
parameters of our NTM trigger theory. For a typical tokamak
plasma, the resistive time is of the order of a few second, hence
the islands saturate in roughly 50-100 ms, while the peak
frequency for the plasma turbulence is around 100 KHz (as observed
for example in Ref.\cite{Romanelli2006}). According to
experimental observation of zonal fields by Fujizawa et al.
\cite{Fujisawa2008}, current corrugations evolve roughly 50 times
slower than the drift-wave turbulence. This is in agreement with
the theoretical prediction that zonal fields are mesoscale
objects. The extrapolation of these results to typical tokamak
plasmas leads to an expected coherence time for the corrugations
around 1-2 ms. As a consequence, in standard conditions, the value
of $\zeta$ is in general below the critical value predicted with
our model, which is in agreement with the fact that, in tokamaks,
seedless NTMs are the exception rather than the rule. However,
current corrugation of sufficient amplitude occurring in
particular plasmas (e.g. where the local resistivity is
particularly high or the zonal fields are slow) could explain the
results in Refs.\cite{Gude1999,Buttery2002,Fredrickson2002}.
Furthermore, assuming that the zonal field time scale will be
roughly unchanged in the next generation tokamaks, such as ITER,
our results seem to suggest that these machines should be stable
to spontaneous NTMs, as their resistive time is of the order of
hundreds of seconds.

A limitation of this work is the assumption that the pressure
profile (and therefore the pressure perturbations) does not play
any role in the calculation of $\Delta'$. A "zonal" pressure is
generated by drift-waves when Finite Larmor Radius effects are
considered and should become significant when ITG instabilities
are present. Pressure gradients might be relevant as they
contribute to Eq.\ref{4} with a term that is inversely
proportional to $(\psi_{eq}')^2$, and are therefore very important
in the neighborhood of the resonant surface. As pointed out in
Refs.\cite{Bishop1991} this could make even harder to properly
evaluate the stability parameter. The inclusion of this effect in
our calculation will be discussed in a future publication.

The authors acknowledge fruitful discussions with Dr. J.W. Connor,
Prof. S.C. Cowley and Dr. A. Thyagaraja. This work was funded by
the United Kingdom Engineering and Physical Sciences Research
Council and by the European Communities under the contract of
Association between EURATOM and UKAEA. The views and opinions
expressed herein do not necessarily reflect those of the European
Commission.

\end{document}